\documentstyle[11pt,twoside,psfig,jltp]{article}

\title{Some remarks on vortex matter in high-$T_{c}$ superconductors}

\author{I. L. Landau \address{Laboratorium f\"ur 
Festk\"orperphysik, ETH H\"onggerberg, CH-8093 Z\"urich, 
Switzerland} and H. R. Ott }

\runninghead{I. L. Landau and H. R. Ott}{Vortex matter in high-$T_{c}$ 
            superconductors}
       
\begin{document}

\begin{abstract}

We show that some experimentally observed features of vortex matter in 
high-$T_{c}$ superconductors may be interpreted in simpler ways than it 
is usually done. In particular, we consider magnetic flux creep at low 
temperatures as well as the irreversibility line in the $H-T$ phase diagram. 
We also discuss a new approach to the analysis of the equilibrium 
magnetization in the mixed state of type-II superconductors and we suggest 
an alternative configuration for the mixed state in magnetic fields close 
to the upper critical field.

\end{abstract}

\maketitle

%Include this space if you do not use sections in your document.
%\vspace{0.3in}

\section{INTRODUCTION}

The behavior of vortex matter in high-$T_{c}$ superconductors (HTSC's) 
seems to be extremely complex. Many different theoretical models have been 
developed in order to explain various experimental observations. The 
complexity of the description of the mixed state in HTSC's and its dynamics 
often arises from the apparent inability of traditional approaches for 
explaining the experimental results. Recent reconsiderations of available 
data indicate, however, that this may not always be the case. In this brief 
review we discuss several features of the mixed state in HTSC's and show 
that in some cases more traditional and less complex models may perfectly 
well explain the experimental observations. This is particularly true for 
the vortex dynamics. We show that thermally activated flux creep as well 
as very specific features of the vortex dynamics, which are usually 
related to the irreversibility line and a vortex-glass transition, may 
quite well be explained by employing a simple Kim-Anderson approach to the 
flux-creep process \cite{1,2} if a profile of the pinning potential 
well is taken into account. We also consider a new approach to the 
analysis of the reversible magnetization $M$ in an external magnetic field 
$H$ and show that the temperature dependence of the upper critical field 
$H_{c2}$ as well as the value of superconducting critical temperature 
$T_{c}$ may reliably be obtained by scaling the $M(H)$ curves measured 
at different temperatures without assuming any specific $M(H)$ dependence. 
Finally, we consider an alternative model for describing the mixed state 
of type-II superconductors in magnetic fields close to $H_{c2}$.

Most of the results that are presented and discussed below are 
included in Refs. \citen{3,4,5,6,7,8}.

\section{VORTEX DYNAMICS}

Our analysis of the vortex dynamics is based on the assumption of 
single vortex hopping. This is not a commonly accepted approach to 
this problem. Nevertheless, the analysis of experimental results 
presented in subsection 2.1 provides rather strong evidence that the 
magnetic relaxation in HTSC's may indeed quite well be described by 
assuming the motion of single vortex lines. \cite{5} We employ here the 
Kim-Anderson approach to the flux motion with the only difference that 
instead of triangular potential barriers, which were implicitly assumed 
by Kim and Anderson in their original work, \cite{2} we consider a more 
realistic, smooth profile of the pinning potential well, as shown by the 
solid line in Fig. 1(a). 
%%%%%%%%%%%
\begin{figure}[h]
\centerline{\psfig{file=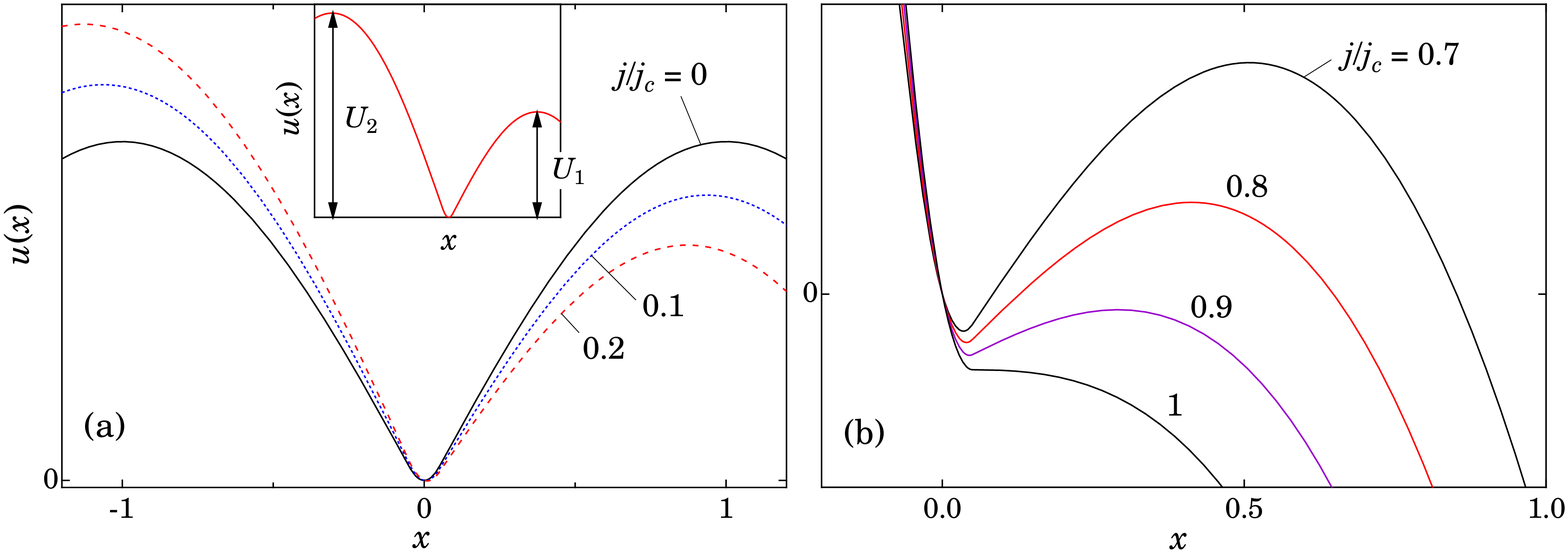,height=1.8 in}}
\caption{Schematic profiles of the pinning well for different values 
         of $j/j_{c}$. (a) $j/j_{c} \ll 1$. (b) $(1 - j/j_{c}) \ll 
         1$. The inset illustrates the meaning  of $U_{1}$ and $U_{2}$. 
         This simple functional $x$ depoendence is chosen only for 
         illustration and has no real physical meaning.}
\end{figure}
%%%%%%%%%%

If the current density $j$ in a superconducting sample in the mixed state 
is less than its critical value $j_{c}$, all vortices are pinned and their 
motion is entirely due to thermally activated hopping of the vortex lines 
or their quantum tunneling. At a given temperature $T$, the profile of 
the potential well for $j = 0$ may be written as 
%%%%%%%
\begin{equation}
u(x)=U(T)f(x,T),
\end{equation}
%%%%%%%
where $U(T)$ is the energy fixing the pinning strength, and the function 
$f(x,T)$ defines the shape of the potential well, which may be temperature 
dependent as well. 

An electric current creates a Lorentz force $F_{L}$ acting on the vortices. 
The Lorentz force tilts the potential profile which facilitates the vortex 
motion in one direction. In the presence of a current the potential profile 
may be written as
%%%%%%%
\begin{equation}
u(x,j)=u(x)-xF_{L}
\end{equation}
%%%%%%%
with $F_{L}= j \delta \Phi_{0}/c$, where $\delta$ is the sample thickness, 
$\Phi_{0}$ is the magnetic flux quantum, and $c$ is the speed of light. 
The variation of the potential profile with increasing current density 
is illustrated in Figs. 1(a) and 1(b). It may be seen that at low currents 
($j \ll j_{c}$) the decrease of the activation energy with increasing 
current is entirely determined by the behavior of $ u(x)$ near its 
maxima, while for currents close to $j_{c}$, only $u(x)$ in the vicinity 
of the inflection point, i.e., where $d^{2}u/dx^{2} = 0$, is important. 
The critical current density is reached if the potential barriers in the 
direction of the vortex motion vanish. According to Eqs. (1) and (2), this 
results in 
%%%%%%%
\begin{equation}
j_{c}=\frac{cU(T)f'_{\max}}{\delta \Phi_{0}},
\end{equation}
%%%%%%%
where $f'_{\max}$ is the value of $df/dx$ at the inflection point. Equation 
(3) represents a formal definition of the critical current density in the 
mixed state of type-II superconductors.

As may clearly be seen in Figs. 1(a) and 1(b), the distance between the 
bottom of the well and the adjacent potential maximum along the direction 
of the flux motion decreases with increasing current and vanishes at 
$j = j_{c}$. This is a direct consequence of Eq. (2) and is true for any 
smooth profile of the potential well. Consequently the flux-creep activation 
energy is always a non-linear function of the current. This non-linearity of 
$U_{1}(j)$ results in a negative curvature of $\ln E$ versus $j$ and in 
a positive curvature of the logarithmic time dependence of the irreversible 
magnetic moment $M_{irr}$. Although these features of the flux-creep  
process, as has been pointed out by Beasley et al., \cite{9} are a 
direct consequence of the Kim-Anderson approach, deviations of 
$M_{irr}(\ln t)$ from linearity are often considered in the literature 
as being incompatible with the Kim-Anderson model.

\subsection{Thermally activated flux creep at low temperatures}

The experiments that we describe in this section have been carried out 
on a ring-shaped YBa$_{2}$Cu$_{3}$O$_{7-x}$ (YBCO) film with a thickness 
of 0.3 $\mu$m. The external diameter of the ring was 10 mm and its width 
was approximately 2 mm. The current in the ring $I$ and its decay due to 
flux creep were measured with a Hall probe placed in the central part of 
the ring cavity (see Refs. \citen{3,4,5} for details).

The main advantage of choosing a ring geometry for this type of experiments 
is the possibility to obtain voltage-current ($V-I$) characteristics 
in the flux creep  regime. Both the analysis and the interpretation of the 
$V-I$ curves are much less influenced by employing different models 
than is usually the case for the analysis of experimental curves of magnetic 
relaxation. The voltage around the ring can straightforwardly be evaluated 
from the experimentally measured $I(t)$ data via $V = L  dI/dt$, where 
$L$ is the sample inductance. The second advantage is that almost all the 
magnetic flux is concentrated inside the ring cavity, while the current 
flows around it. In this case, the non-uniformity of the current 
distribution in the sample's cross-section can be neglected and all 
complications arising from the use of critical-state models are avoided. 
Examples of experimental voltage-current characteristics plotted as $T 
\ln V$ versus $I$ are shown in Fig. 2.
%%%%%%%%%%%
\begin{figure}[h]
\centerline{\psfig{file=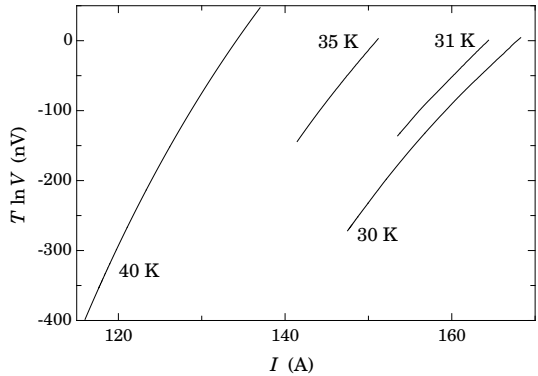,height=2 in}}
\caption{Examples of voltage-current characteristics of a 
         ring-shaped YBCO film at $H=0$.}  
\end{figure}
%%%%%%%%%%

At temperatures not very close to $T_{c}$, the probability of vortex 
hopping is negligible at low currents and therefore, the flux motion can 
be observed only at current densities sufficiently close to $j_{c}$. In 
this case, the vortex hopping in the direction opposite to the Lorentz 
force can be neglected and the hopping rate $\nu$ is entirely determined by 
the height of the potential barrier $U_{1}$ in the direction of the Lorentz 
force (see inset to Fig. 1(a)), i.e.,
%%%%%%%
\begin{equation}
\nu=\nu_{0} \exp (-U_{1}/k_{B}T),
\end{equation}
%%%%%%%
where $\nu_{0}$ is the attempt frequency, $k_{B}$ is the Boltzmann 
constant, and $U_{1}$ plays the role of the activation energy in the 
flux-creep process. Taking into account that the electric field $E$ 
is proportional to the hopping rate, we can rewrite Eq. (4) as
%%%%%%%
\begin{equation}
U_{1}(j,T)=- k_{B}T[\ln E(j) - \ln E_{0}],
\end{equation}
%%%%%%%
where $E_{0}$ is a parameter which includes the attempt frequency, the 
hopping distance and the magnetic induction. Because $U_{1}(j_{c},T) = 
0$, $\ln E_{0} = \ln E(j_{c})$.

Eq. (5) provides the possibility to obtain the current dependence 
of the flux-creep activation energy from the experimental $E-j$ 
characteristics. The problem is that the experimental data for each 
temperature cover only a very narrow range of currents and, therefore, 
only a very small part of the $U_{1}(j)$ curve at a given temperature can 
be obtained in this way. An additional complication in this evaluation 
of $U_{1}(j)$ is that neither $E_{0}$ nor $j_{c}$ are {\it a priori} known. 
However, with some additional assumptions about the temperature dependence 
of the pinning potential profile, Eq. (5) is adequate for the analysis 
of flux-creep data obtained at different temperatures. Not only can 
$U_{1}(j)$ curves be determined for a much wider range of currents, but 
also the quantities $E_{0}$ and $j_{c}$ may be evaluated. The problem here 
is to select reasonable assumptions. As will be shown below, the assumption 
that the function $f$ in Eq. (1) is temperature independent is quite 
consistent with experimental data in a very wide range of temperatures. 
In this case, according to Refs. \citen{3} and \citen{5}, the flux-creep 
activation energy may be written as 
%%%%%%%
\begin{equation}
U_{1}(j,T)=U(T)Y(j/j_{c}),
\end{equation}
%%%%%%%
where $U(T)$ is the same as in Eq. (1) and the function $Y$ depends 
only 
on the ratio $j/j_{c}$. 

It has been demonstrated in Refs. \citen{3} and \citen{5} that, if the 
flux-creep activation energy may be represented as a product of a 
temperature dependent and a current dependent term, the transformation 
%%%%%%%
\begin{equation}
\ln E(j/i,T_{0})=(T/iT_{0}) \ln E(j,T)+A
\end{equation}
%%%%%%%
with $i = j_{c}(T)/j_{c}(T_{0})$ and
%%%%%%%
\begin{equation}
A = (1 - T/iT_0) \ln E_{0}
\end{equation}
%%%%%%%
may be used to merge the $\ln E-j$ curves measured at different temperatures 
into a single master curve. Here, $i$ and $A$ are scaling parameters, and 
$T_{0}$ is some arbitrary chosen temperature within the investigated 
temperature range. The resulting master curve represents the current 
dependence of $\ln E$ at $T = T_{0}$, as if $E(j)$ could actually be 
measured over this extended range of currents at this single temperature. 
For each temperature the values of $i$ and $A$ can be found from forcing 
the overlapping $T \ln E$ versus $j$ curves for the adjacent temperatures 
to match each other. In this procedure the relation between $i$ and $A$ 
given by Eq. (8) is not employed, both quantities are rather considered 
as independent fitting  parameters. Eq. (8) is only used retrospectively 
in order to check the validity of our approach. 
%%%%%%%%%%%
\begin{figure}[t]
\centerline{\psfig{file=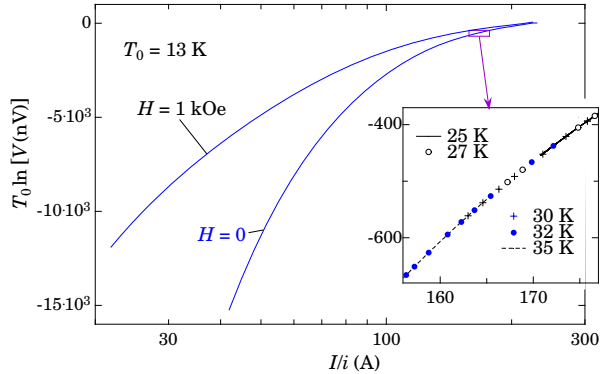,height=2 in}}
\caption{Results of the scaling procedure in the form of $T_{0} \ln 
         V(T_{0})$ versus $I/i$ with $T_{0} = 13$ K. The inset shows, 
         on linear scales, the small part of the curve for $H = 0$ which 
         is indicated by the rectangle in the main figure. For clarity 
         only very few points for each temperature are displayed.}  
\end{figure}
%%%%%%%%%%

In the following discussion of experimental results we use measurable 
quantities, such as the voltage $V$ and the current $I$, but will switch 
to $E$ and $j$ in relevant equations.

The feasibility of the scaling procedure, implicit in Eq. (7), has been 
demonstrated in Refs. \citen{3} and \citen{5}. In Fig. 3 we show again 
two examples of this procedure, obtained for two different values of 
external magnetic fields. The scaling procedure provides the corresponding 
master curves, exhibiting a practically perfect alignment of the $T \ln 
V$ versus $I$ curves measured at different temperatures between 10 and 
80 K. The inset of Fig. 3 emphasizes the matching quality on extended 
scales. 

We note that the behavior of the $\ln V$ versus $I$ curves changes 
drastically at temperatures $T \le 10$ K and the low temperature data cannot 
be scaled using Eq. (7). \cite{3,4} This is a clear indication for a 
crossover from the thermally activated vortex hopping to quantum tunneling 
of the vortex lines with decreasing temperature. As was shown in Ref. 
\citen{4}, our scaling procedure can easily be modified for the analysis 
of the quantum creep regime.

The resulting temperature dependence of the scaling parameter $i$ is shown 
in Fig. 4(a). This plot represents the temperature dependence of the 
normalized critical current, exhibiting the expected trend to saturation 
at low temperatures. Note that, according to Eq. (3) and our 
assumption that $f$ is temperature independent, $j_{c}(T)/j_{c}(T_{0}) = 
U(T)/U(T_{0})$.
%%%%%%%%%%%
\begin{figure}[ht]
\centerline{\psfig{file=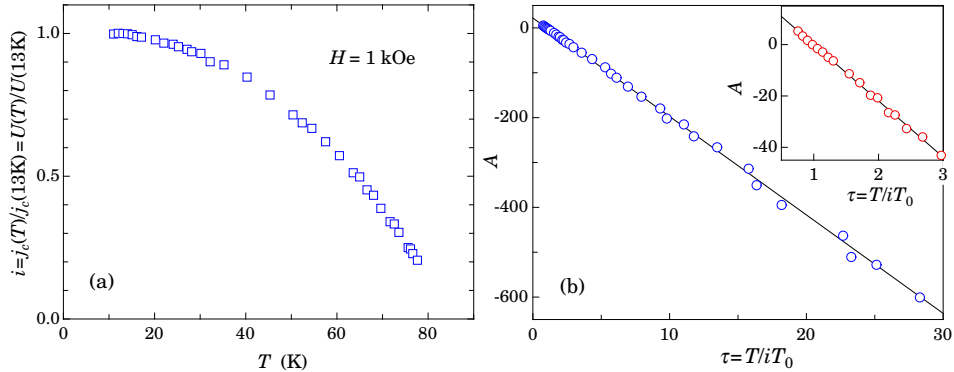,height=2 in}}
\caption{(a) The scaling parameter $i = j_{c}(T)/j_{c}(T_{0})$ as a function 
         of temperature. (b) The parameter $A$ as a function of $\tau = 
         T/iT_{0}$ with $T_{0} = 13$ K. The straight line is the best linear 
         fit to $A(\tau)$  for $\tau < 3$. The inset shows the low 
         temperature part of the plot on expanded scales.}  
\end{figure}
%%%%%%%%%%

In Fig. 4(b) we display the temperature dependence of the scaling parameter 
$A$ which, according to Eq. (8), depends on the ratio $T/i$ rather than 
the temperature alone. Hence $A$ is plotted as a function of $\tau = 
T/iT_{0}$. If the temperature dependence of $\ln V_{0}$ is negligible and 
our procedure makes sense, we expect the data to lie along a straight line. 
Although $V_{0}$ is proportional to the temperature-dependent attempt 
frequency, it enters Eq. (8) only as $\ln V_{0}$ and therefore, the resulting 
curve is expected to deviate only weakly from linearity, as is indeed the 
case. At low temperatures, where the temperature dependence of $\ln 
V_{0}$ may definitely be neglected, the $A(\tau)$ points are indeed well 
approximated by a straight line with a slope $dA/d \tau = - A(0)$, in 
complete agreement with Eq. (8) and convincingly documented in the inset 
of Fig. 4(b). Although the temperature dependence of $A$ itself does not 
carry much of physical information, this perfect agreement between $A(\tau)$ 
obtained from the scaling procedure and $A$ given by Eq. (8) serves as an 
important confirmation of the validity of our approach.  As is discussed 
in more detail in Ref. \citen{5}, the linearity of $A(\tau)$ breaks down at 
temperatures $T > 0.9T_{c} \approx$ 80 K. Our main assumption about the 
temperature independence of the function $f$ entering Eq. (1) is thus not 
valid at temperatures close to $T_{c}$. This is to be expected because 
both the magnetic field penetration depth $\lambda (T)$ and the coherence 
length $\xi (T)$ diverge at $T_{c}$.

From the $A(\tau)$ data, a reliable value of $\ln V_{0}$ may be 
obtained. According to Eq. (8), $\ln V_{0} = dA/d \tau = -A(0)$, and 
thus  $\ln [V_{0}$ (nV)] = 21.9. With $\ln V_{0}$ known, Eq. (5) can 
now be used to established the current dependence of the flux-creep 
activation energy from the data presented in Fig. 3.  

As mentioned above, at current densities close to $j_{c}$, only a small 
part of the $u(x)$ function in the vicinity of the inflection point 
represents the essential part of the potential barrier (see Fig. 1(b)). 
In this case, $u(x)$ can be replaced by its Taylor series expansion. Taking 
into account that at the inflection point $d^{2}u/dx^{2} = 0$ and keeping 
only the first two nonzero terms in the expansion of $u(x)$, one obtains 
\cite{3,9}
%%%%%%%
\begin{equation}
U(j/j_{c}) \propto (1 - j/j_{c})^{3/2}.
\end{equation}
%%%%%%%
This is the current dependence of the flux-creep activation energy for 
$(1 - j/j_{c}) \ll 1$ which does not depend on the particular shape of 
$u(x)$. Eq. (9), together with Eq. (5) serves to estimate the value 
of the critical current and we obtain $I_{c}$(1 kOe) = 300 A at $T = 
13$ K. \cite{5} Taking into account that $I_{c}$ is almost constant at 
these low temperatures, this value of  $I_{c}$ may safely be considered 
as the critical current for $T = 0$. 
%%%%%%%%%%%
\begin{figure}[h]
\centerline{\psfig{file=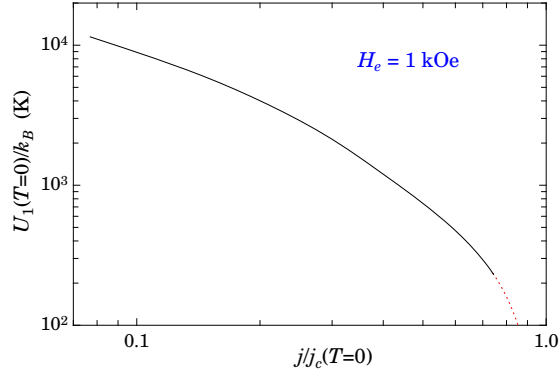,height=2 in}}
\caption{$U_{1}(j/j_{c})$  calculated for $T = 0$ is shown by the solid 
         line. The dotted line is extrapolation of the $U_{1}(j/j_{c})$ 
         curve to $j/j_{c} = 1$ using Eq. (9).}  
\end{figure}
%%%%%%%%%%

We may now insert the values $I_{c}$ and $\ln V_{0}$ into Eq. (5) and 
calculate $U_{1}(j/j_{c})$ from the master curves presented in Fig. 
3. The result is shown in Fig. 5. The extrapolation of the 
$U_{1}(j/j_{c})$ curve to $j/j_{c} = 1$ using Eq. (9) is shown by the 
dotted line.

Our approach assumes a direct connection between the profile of the 
potential well $u(x)$ and $U_{1}(j/j_{c})$, which allows for the
reconstruction of the potential profile in real space using the 
$U_{1}(j/j_{c})$ data as they follow from experiment. \cite{3,5} The result 
is shown in Fig. 6(a). Here, for simplicity, we have chosen that the 
inflection point of the $u(x)$ function coincides with the bottom of the 
potential well. The evolution of the potential profile with increasing 
current is shown in Fig. 6(b). As may be seen in Fig. 5, our 
$U_{1}(j/j_{c})$ data are limited to currents $j \ge 0.08 j_{c}$. This 
is why only a limited section of the potential profile for $\left| x \right| < 
x_{\max} \approx$ 67 {\AA} can be obtained. No information about the potential 
profile at distances larger than $x_{\max}$ can be gained on the basis of 
our experimental results. 
%%%%%%%%%%%
\begin{figure}[t]
\centerline{\psfig{file=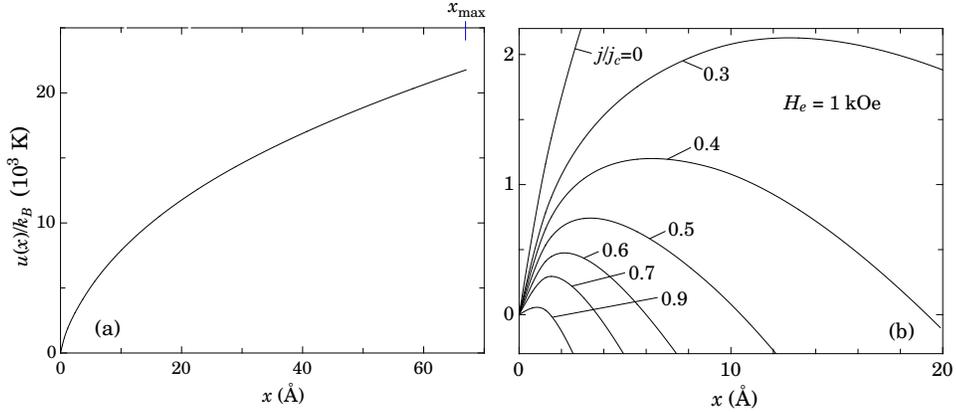,height=2.2 in}}
\caption{(a) The pinning potential profile for $j = 0$ calculated from 
         $U_{1}(j/j_{c})$. (b) The potential profile for different 
         current densities.}  
\end{figure}
%%%%%%%%%%

Here we have used the current dependence of the flux-creep activation 
energy to calculate the profile of potential barriers. On the other hand, 
it is well known that HTSC samples are not uniform and one should expect 
that different barriers have different shapes. Hence the physical relevance 
of the potential profiles calculated as demonstrated above, is not obvious. 
In order to clarify the situation, we consider the flux-creep process in 
more detail. There are very many different trajectories along which the 
vortices are allowed to cross the ring sample. It is obvious, however, that 
only those trajectories containing the lowest potential barriers will actually 
be traced. Along each trajectory many different potential barriers are 
met, but only one or a very few of them with the largest amplitudes are 
essential. The next question to be answered is, how many trajectories are 
needed to let all the vortices pass across the sample. In the ring geometry, 
the evaluation of the number $N$ of vortices which are leaving or entering 
the ring cavity per second is straightforward. Taking into account that the 
experimentally accessible voltages range between $3\cdot 10^{-5}$ and 0.3 nV, 
$N$ is between 20 and $2\cdot 10^{5}$ s\textsuperscript{-1} for the lowest 
and the highest voltage, respectively. For $B = 1$ kG, the distance between 
vortices is of the order of $10^{-5}$ cm. This implies an average vortex 
velocity $w \approx 2$ cm/s, if we force all the $2\cdot 10^{5}$ vortices per 
second to follow the same trajectory across the sample. This value of 
$w$ is rather low and there is good reason to assume that a single trajectory 
is in principle sufficient to transfer all the vortices.

An important consequence of this line of thoughts is that the analysis 
of flux-creep rates provides information only about one particular pinning 
center, which represents the highest potential barrier for the vortex motion 
on the energetically most favorable trajectory across the sample. This 
is true not only for the ring geometry, but for measurements of the 
magnetization relaxation for other sample shapes, as well. 

Although it is claimed by many that single vortex hopping cannot 
describe experimental observations reflecting flux-creep in HTSC's, we 
have demonstrated that this is not really the case. The described 
procedure for the analysis of flux-creep rates measured at different 
temperatures allows for at least a partial reconstruction of the pinning 
potential well profile in real space. All the flux-creep data in the 
temperature range between 10 and 80 K, where our scaling procedure is 
applicable, may perfectly well be described by a reasonable potential profile 
shown in Fig. 6(a), with $U(T)/U(0)$ as shown in Fig. 4. 

\subsection{Irreversibility line}

The irreversibility line (IRL) in the $H-T$ phase diagram separates two 
regions with distinctly different behaviors. Above the IRL the magnetization 
of the sample is perfectly reversible, i.e., the sample cannot carry any 
persistent current. Only below the IRL, irreversible magnetization 
$M_{irr}$ is observed. \cite{10,11,12,13,14,15,16,17}
%%%%%%%%%%%
\begin{figure}[t]
\centerline{\psfig{file=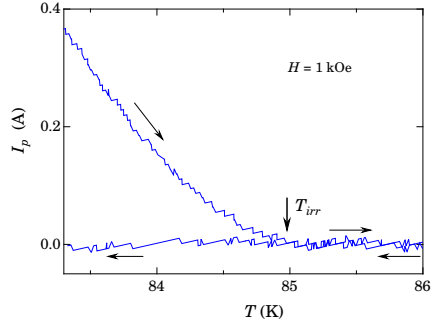,height=2 in}}
\caption{Persistent current $I_{p}$ in the ring-shaped 
        YBa$_{2}$Cu$_{3}$O$_{7-x}$ film as a function of temperature. The 
        vertical arrow indicates the position of the irreversibility 
        temperature. The sample was cooled in an external magnetic field 
        $H = 1$ kOe to $T = 82$ K. The current was subsequently induced 
        by enhancing the magnetic field by 1 Oe.}  
\end{figure}
%%%%%%%%%%

The situation arising near the IRL may much easier be analyzed if we 
again consider a ring-shaped sample. In this case, instead of magnetization  
curves, the temperature variations of the persistent current $I_{p}$ may 
be considered. Typical experimental data for our ring-shaped YBCO film 
are presented in Fig. 7, displaying a heating-cooling cycle of $I_{p}$. 
These data as well as results of measurements of the irreversible 
magnetization, \cite{10,11,12,13,14,15,16,17} reveal that above the 
irreversibility temperature, $T_{irr}$, persistent currents are essentially 
zero. For this reason, $T_{irr}$ is usually considered as the temperature 
at which the critical current density vanishes. It is commonly accepted 
that the melting of the vortex-glass is responsible for such a behavior. 
\cite{18,19} However, as we demonstrate below, this type of $I_{p}(T)$ 
or $M_{irr}(T)$ curves necessarily follows from the simplest Kim-Anderson 
approach for describing the thermally activated vortex motion and the 
critical current density does not really vanish at $T = T_{irr}$ but 
remains nonzero also above the IRL.

Encouraged by the success of the Kim-Anderson approach in the analysis 
of low-temperature flux-creep rates, we now consider the same concept at 
temperatures close to $T_{c}$. In this case, the vortex hopping in the 
direction opposite to the Lorentz force cannot be neglected and the 
electric field in the sample is 
%%%%%%%
\begin{equation}
E=E_0\left\{ {\exp \left[ {-{{U_1(T,j)} \over {k_BT}}} \right]-\exp 
\left[ 
{-{{U_2(T,j)} \over {k_BT}}} \right]} \right\}.
\end{equation}
%%%%%%%
The second term in Eq. (10) describes the vortex hopping in the 
direction opposite to that of the the Lorentz force (see Fig. 1(a) 
for the definitions of $U_1$ and $U_2$). We note that, because $U_{1}$ 
and $U_{2}$ depend on current differently, Eq. (10) cannot be reduced 
to a hyperbolic sinus.

The electric field $E$ is proportional to the current decay rate $dj/dt$ 
and therefore Eq. (10) may be used for its evaluation. In order to 
calculate the temperature and current dependencies of $dj/dt$, we have 
to assume some explicit expression for the profile of the potential well 
$u(x)$. For the following analysis we have chosen two rather different 
representations for $f(x,T)$, i.e.,
%%%%%%%
\begin{equation}
f(x,T)=\left| x \right|-(1-T/T_c)^kx^2
\end{equation}
%%%%%%%
and
%%%%%%%
\begin{equation}
f(x,T)=\left( {\sqrt {\left| x \right|+x_0}-\sqrt {x_0}} 
\right)-a{{\left( 
{\left| x \right|+ x_0} \right)^{3/2}-x_0^{3/2}} \over {\left( 
{1-T/T_c} \right)^m}},
\end{equation}
%%%%%%%
with
%%%%%%%
\begin{equation}
U(T)=U_{0}(1-T/T_{c})^{3/2}
\end{equation}
%%%%%%%
for both cases. According to Eq. (1) the profile of the potential well 
$u(x,T) = U(T)f(x,T)$. Our choice of $f$ functions, shown in Fig. 8, is 
quite arbitrary and was mainly dictated by the possibility of performing 
analytical calculations. As will be shown below, the particular choice 
of $f(x,T)$ does not influence the main qualitative features of the 
flux-creep process. 
%%%%%%%%%%%
\begin{figure}[h]
\centerline{\psfig{file=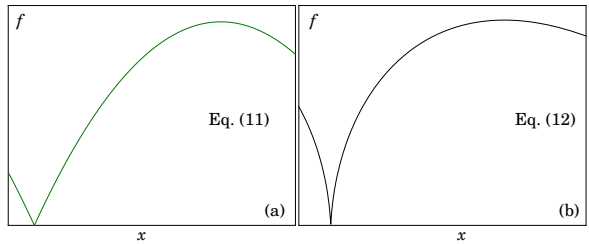,height=1.3 in}}
\caption{(a) and (b) Examples of profiles of the pinning potential wells 
         as given by Eqs. (11) and (12).}  
\end{figure}
%%%%%%%%%%

With the chosen potential profiles, the dependencies of $dj/dt \propto 
E$ on current and temperature may straightforwardly be calculated using 
Eqs. (1), (2) and (10). Fig. 9(a) displays the results of calculations 
of $dj/dt$ for $f(x,T)$ given by Eq. (11) with $k = 1.4$ at several 
fixed current densities. We use a log-scale for the $dj/dt$-axis and the 
total change in $dj/dt$ is 100 orders of magnitude. This figure clearly 
demonstrates that $dj/dt$ grows extremely fast with increasing temperature 
which is exactly the experimentally observed behavior of the current decay 
near the irreversibility temperature. Note that all the data presented 
in Fig. 9(a) correspond to $j < j_{c}$.
%%%%%%%%%%%
\begin{figure}[h]
\centerline{\psfig{file=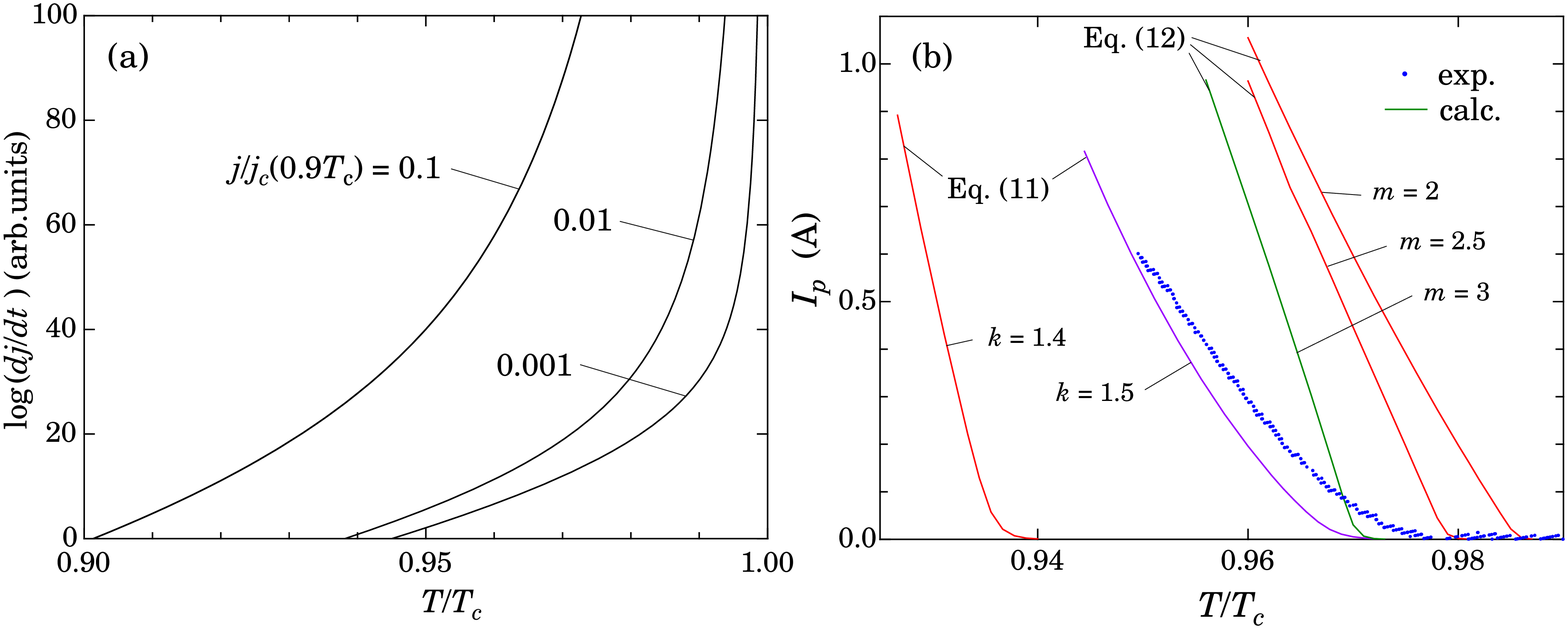,height=2 in}}
\caption{(a) $dj/dt$ versus $T/T_{c}$ calculated for $f(x,T)$ given by 
         Eq. (11) with $k = 1.4$. The values of the current density normalized 
         by $j_{c}$ at $T = 0.9T_{c}$ are indicated for each curve. 
         (b) The persistent current as a function of temperature. The solid 
         lines are the results of calculations for different representations 
         of $u(x)$. The points are experimental data from Fig. 7.}  
\end{figure}
%%%%%%%%%%

Experimentally the persistent current $I_{p}$ is usually determined as 
the current that does not decay during the time of the experiment. With our 
approach we can calculate the temperature dependence of the persistent 
current by fixing $dI/dt$ according to experimental conditions. The 
corresponding calculations were made for $dI/dt = 3\cdot 10^{-4}$ A/s, 
which is close to the resolution of the experimental data presented in 
Fig. 7 (see Ref. \citen{6} for details). In order to understand how the 
$I_{p}(T)$ curves depend on the particular choice of the potential profile, 
the calculations were made for both approximations of $u(x)$ and for 
different values of the exponents $k$ and $m$ in Eqs. (11) and (12), 
respectively. The results of the calculations, together with experimental 
data from Fig. 7, are shown in Fig. 9(b). It may be see that an 
"irreversibility" temperature exists for all chosen $u(x)$ functions. At 
the same time, both the shape of the $I_{p}(T)$ curves and the position 
of the apparent $T_{irr}$ are rather sensitive to the choice of $u(x)$. 
This means that by a proper choice of $U(T)$ and $f(x,T)$ entering Eq. 
(1), any experimentally observed temperature dependence of the irreversible 
magnetic moment or the persistent current may sufficiently well be approximated 
and no specific transition in the vortex system is needed to explain the 
existence of the experimentally observable irreversibility line. 

\subsection{Vortex-glass transition}

The vortex-glass transition is undoubtedly the most popular interpretation 
of the IRL in HTSC's. Many experimental results, especially the scaling 
of $\log E - \log j$ curves measured at different temperatures, seem to 
confirm the concept of a vortex-glass melting at temperatures close to 
the IRL. \cite{14,20,21,22,23,24,25,26,27,28,29} Quite recently, however, 
it has been shown that the same experimental $\log E - \log j$ data may 
be scaled equally well by assuming very different temperatures of the 
vortex-glass transition. \cite{30} This observation is important because 
it demonstrates very clearly that the concept of vortex-glass melting, 
although commonly accepted, is not actually confirmed by experiment. In 
this section we show that the distinct variation of the shape of the $\log 
E - \log j$ curves, which is usually interpreted as a manifestation of 
the vortex-glass melting, follows straightforwardly from Eq. (10). 
%%%%%%%%%%%
\begin{figure}[h]
\centerline{\psfig{file=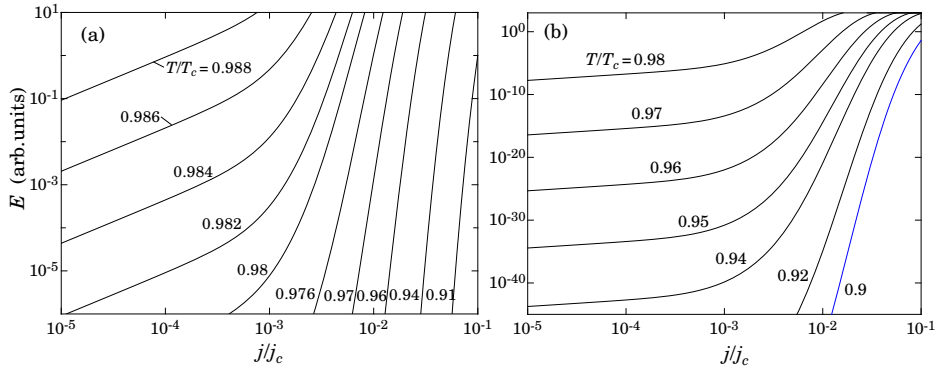,height=1.95 in}}
\caption{$E-j$ curves calculated for $f(x,T)$ given by Eq. (12) 
         with $m = 2$.}  
\end{figure}
%%%%%%%%%%

For the calculations, whose results are presented in this section, we used 
$f(x,T)$, as given by Eq. (12) with $m = 2$. The results of the calculations 
of $E(j)$ are shown in Figs. 10(a) and 10(b) as $E(j)$ curves at fixed 
temperatures. Qualitatively the plot shown in Fig. 10(a) is indistinguishable 
from numerous experimental results (see Refs.  
\citen{20,21,22,23,24,25,26,27,28,29}) and it is clear that, by corresponding 
adjustments of $U(T)$ and $f(x,T)$, any experimental $E(j,T)$ curve may 
be approximated even quantitatively. We note that the calculated $\log 
E - \log j$ curves remain qualitatively the same for any chosen potential 
profile. 

Fig. 10(b) shows the $E(j)$ curves down to much lower voltages and, as 
may clearly be seen, the change of the sign of the curvature, which is 
usually attributed to the vortex-glass transition, is a universal feature 
of the $E(j)$ curves at any temperature. With decreasing temperature however, 
the sign change of the curvature is shifted to a range of voltages 
which is not accessible experimentally. 

A very similar explanation of the voltage-current characteristics near 
the "vortex-glass" transition has been suggested by Coppersmith at 
al.. \cite{31} In their short comment they considered a sinusoidal potential 
barrier. It was shown that even with this simple potential all the qualitative 
features of the experimental $E(j)$ curves could be reproduced. The authors 
also pointed out that the insignificant quantitative disagreement with 
the experimental data is simply due to the arbitrary chosen sinusoidal 
profile of the potential barriers. 

\section{TEMPERATURE DEPENDENCE OF $H_{c2}$ FROM ISOTHERMAL 
MAGNETIZATION DATA.}

In order to evaluate the upper critical field, $H_{c2}$, from experimental 
data in complex materials such as HTSC's, it is very important to introduce 
an appropriate definition of this parameter. In an ideal type-II 
superconductor, $H_{c2}$ is the highest value of the magnetic field compatible 
with superconductivity, i.e., the $H_{c2}(T)$ curve in the $H-T$ phase diagram 
represents a line of second order phase transitions to the normal state. 
As is well known for HTSC's, this transition degenerates to a cross-over 
region because of fluctuation effects. We note that small inclusions of 
impurity phases with critical temperatures different from that of the bulk 
cannot always be excluded in HTSC's and they may also contribute to the 
broadening of the transition. At the same time, in magnetic fields well below 
$H_{c2}(T)$, the effect of fluctuations and possible inclusions of impurity 
phases on the sample magnetization is small and the $M(H)$ curves in this 
magnetic field range must practically be the same as for a perfectly uniform 
sample without fluctuations. This circumstance provides the possibility 
to evaluate the temperature dependence of $H_{c2}$, in its traditional 
sense, from magnetization measurements in magnetic fields well below 
$H_{c2}$.

Here we discuss a new approach to this problem by scaling the $M(H)$ curves 
measured at different temperatures. This scaling procedure is based on 
the application of the  Ginzburg-Landau (GL) theory in very general terms, 
without assuming any specific magnetic field dependence of the 
magnetization. We consider this as an important point because reliable 
calculations of $M(H)$ are extremely difficult even for uniform and isotropic 
superconductors. The reliability of approximate models can in most cases 
not independently be verified and hence their application can easily lead 
to misinterpretations of experimental results. 

The scaling procedure is based on the assumption that the GL parameter 
$\kappa$ is temperature independent. Although the microscopic theory of 
superconductivity predicts a temperature dependence of $\kappa$, 
\cite{32,33} this dependence is rather weak and is not expected to change 
the results significantly. From the GL theory it follows straightforwardly 
that, if $\kappa$ is temperature independent, the magnetic susceptibility 
$\chi$ of the sample is a universal function of $H/H_{c2}$, \cite{34} 
i.e., $\chi (H,T) = \chi (h)$ with $h = H/H_{c2}(T)$, and the magnetization 
density is
%%%%%%%
\begin{equation}
M(H,T)=H_{c2}(T)h \chi (h).
\end{equation}
%%%%%%%
Eq. (14) leads to the following relation between the values of $M$ at two 
different temperatures 
%%%%%%%
\begin{equation}
M(H,T_{0})=M(h_{c2}H,T)/h_{c2}.
\end{equation}
%%%%%%%
with $h_{c2} = H_{c2}(T)/H_{c2}(T_{0})$. The collapse of individual 
$M(H)$ curves measured at different temperatures into a single master 
curve may be achieved by a suitable choice of $h_{c2}(T)$. In this way 
one can only establish the temperature dependence of the normalized 
upper critical field $H_{c2}(T)/H_{c2}(T_{0})$, while the absolute values 
of $H_{c2}(T)$ remain unknown. This is the price to pay for the fact 
that we do not specify the variation of the magnetization upon changing 
the applied magnetic field. 

The scaling procedure described by Eq. (15) is actually valid for ideal 
type-II superconductors only and in order to apply it to HTSC's, we have 
to introduce the necessary corrections to Eq. (15) that are dictated by 
some specific features of HTSC's. Most of the families of HTSC's 
exhibit a weak paramagnetic susceptibility in the normal state. Its 
influence may be accounted for by replacing Eq. (15) by
%%%%%%%
\begin{equation}
M(H,T_{0})=M(h_{c2}H,T)/h_{c2} + c_{0}(T)H.
\end{equation}
%%%%%%%
The term $c_{0}(T)H$ also includes contributions to the sample magnetization 
arising from fluctuations effects and small inclusions of impurity phases 
which, of course, can only approximately be accounted for. However, as 
will be shown below, Eq. (16) can be successfully used for the scaling 
of experimental $M(H)$ data up to temperatures quite close to $T_{c}$. 
In the following we use the parameter $c_{0}(T)$ in Eq. (16) as an additional 
adjustable parameter in the scaling procedure.
%%%%%%%%%%%
\begin{figure}[h]
\centerline{\psfig{file=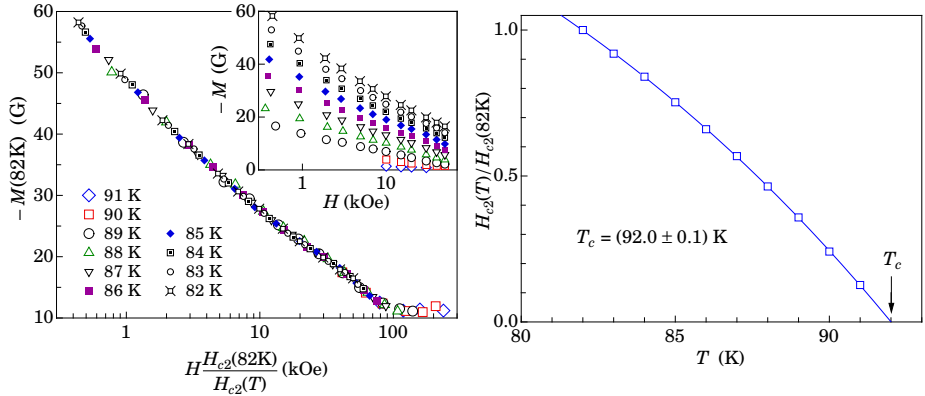,height=2.1 in}}
\caption{(a) The magnetization data for a sample of YBa$_{2}$Cu$_{3}$O$_{7-x}$ 
         (Ref. \protect\citen{35}) after scaling using Eq. (4) with 
         $T_{0} = 82$ K. The inset displays the original data. (b) 
         $H_{c2}(T)/H_{c2}(82K)$. The solid line is the best fit with Eq. 
         (17).}
\end{figure}
%%%%%%%%%%

The result of this scaling procedure for an Y-based cuprate sample is shown 
in Fig. 11(a). It may be seen that a rather perfect overlap of the individual 
$M(H)$ curves, measured at different temperatures, is obtained in this way. 
Because the variable $h_{c2}$ enters the denominator of the first term 
in Eq. (16), the magnetization data for the highest temperatures are 
considerably expanded in comparison with the low temperature data. This is 
the reason for the somewhat enhanced scatter in the high temperature data. 
The resulting temperature dependence of the normalized upper critical field 
for this sample is shown in Fig. 11(b). 

We note that the uncertainty of $h_{c2}(T)$ increases considerably for 
temperatures close to $T_{c}$ as well as for the lowest temperatures. The 
loss of accuracy for the highest temperatures is due to the obvious 
enhancement of the experimental uncertainty of the original $M(H)$ data. 
Although the experimental accuracy is improving with decreasing temperature, 
the increase of the irreversibility field limits the available magnetic 
field range. If the experimental data are only collected in a narrow magnetic 
field range, our scaling procedure is not reliable.
%%%%%%%%%%%
\begin{figure}[t]
\centerline{\psfig{file=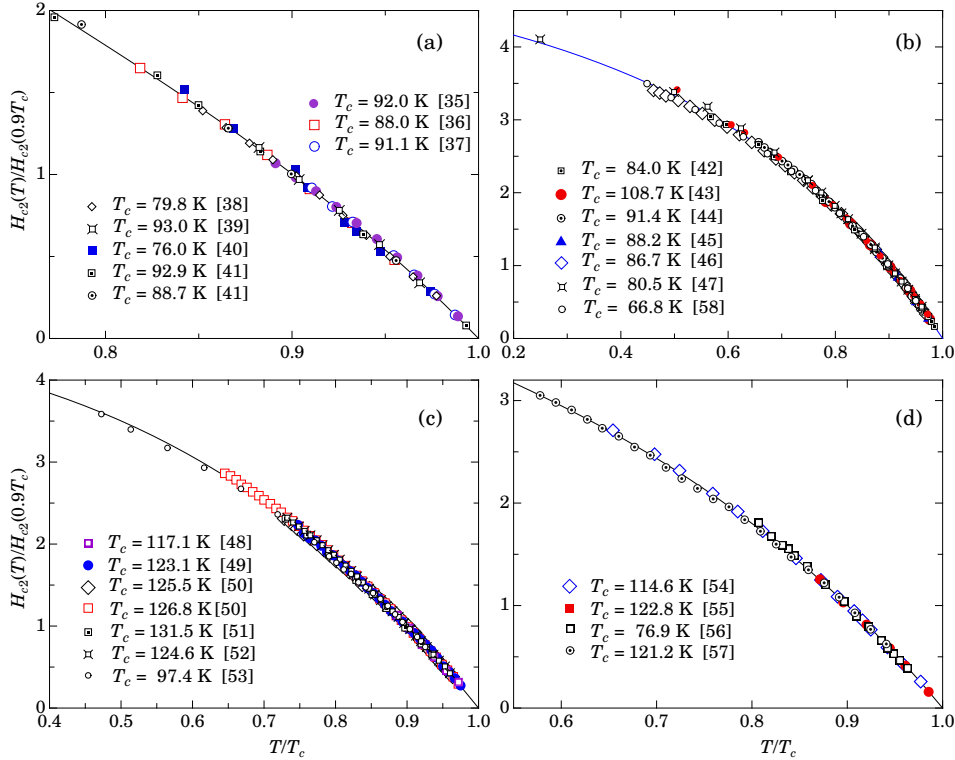,height=4 in}}
\caption{$H_{c2}(T)/H_{c2}(0.9T_{c})$ versus $T/T_{c}$ for different samples; 
         the values of $T_{c}$ are estimated by extrapolating the 
         $h_{c2}(T)$ curves and the corresponding references are 
         indicated near the symbols. The solid lines are guides to the 
         eye. (a) Y-based cuprates. (b) Bi-based cuprates. (c) Hg-based 
         cuprates. (d) Tl-based cuprates.}  
\end{figure}
%%%%%%%%%%

The temperature dependence of the normalized upper critical field, as shown 
in Fig. 11(b), may also be used to evaluate the critical temperature 
$T_{c}$. For this purpose the ratio $H_{c2}(T)/H_{c2}(T_{0})$ was 
approximated by 
%%%%%%%
\begin{equation}
\frac{H_{c2}(T)}{H_{c2}(T_{0})}=\frac{1-(T/T_{c})^{\mu}}{1-(T_{0}/T_{c})^{\mu}},
\end{equation}
%%%%%%%
in which $\mu$ and $T_{c}$ are used as fit parameters. Eq. (17) provides 
a rather good approximation to $h_{c2}(T)$ curves for $T \ge 0.8T_{c}$. 
The corresponding fit is shown as the solid line in Fig. 11(b). The resulting 
value of $T_{c}$ is indicated in Fig. 11(b). If the experimental data 
are obtained up to temperatures close to the critical temperature, the  
value of $T_{c}$, estimated by the extrapolation of the $h_{c2}(T)$ curve 
to $h_{c2} = 0$ is quite accurate. A reliable value of $T_{c}$ is essential 
for the comparison of the results that were obtained for samples with 
different critical temperatures. Using the values of $T_{c}$ evaluated 
in such a way, we have plotted $H_{c2}(T)/H_{c2}(0.9T_{c})$ versus 
$T/T_{c}$ for various Y-based compounds as shown in Fig. 12(a). Quite 
surprisingly, the temperature variations of $H_{c2}$ for rather different 
samples turn out to be identical. 

The temperature variations of $h_{c2}$ for other families of HTSC's are 
plotted in Figs. 12(b-d). Similar to what has been found for Y-based 
compounds, the scaling procedure again leads to an almost perfect merging 
of all the data into one single curve for different samples. Furthermore, 
as may clearly be seen in Fig. 13, the temperature dependencies of the 
normalized upper critical field for different families of HTSC are virtually 
identical at all temperatures for which the experimental data are available. 
\cite{65a} The insignificant differences between the $h_{c2}(T/T_{c})$ 
curves for different samples, visible at the lowest temperatures in Figs. 
12(b), 12(c), and 13, are due to small errors in the evaluation of the 
critical temperature. 
%%%%%%%%%%%
\begin{figure}[h]
\centerline{\psfig{file=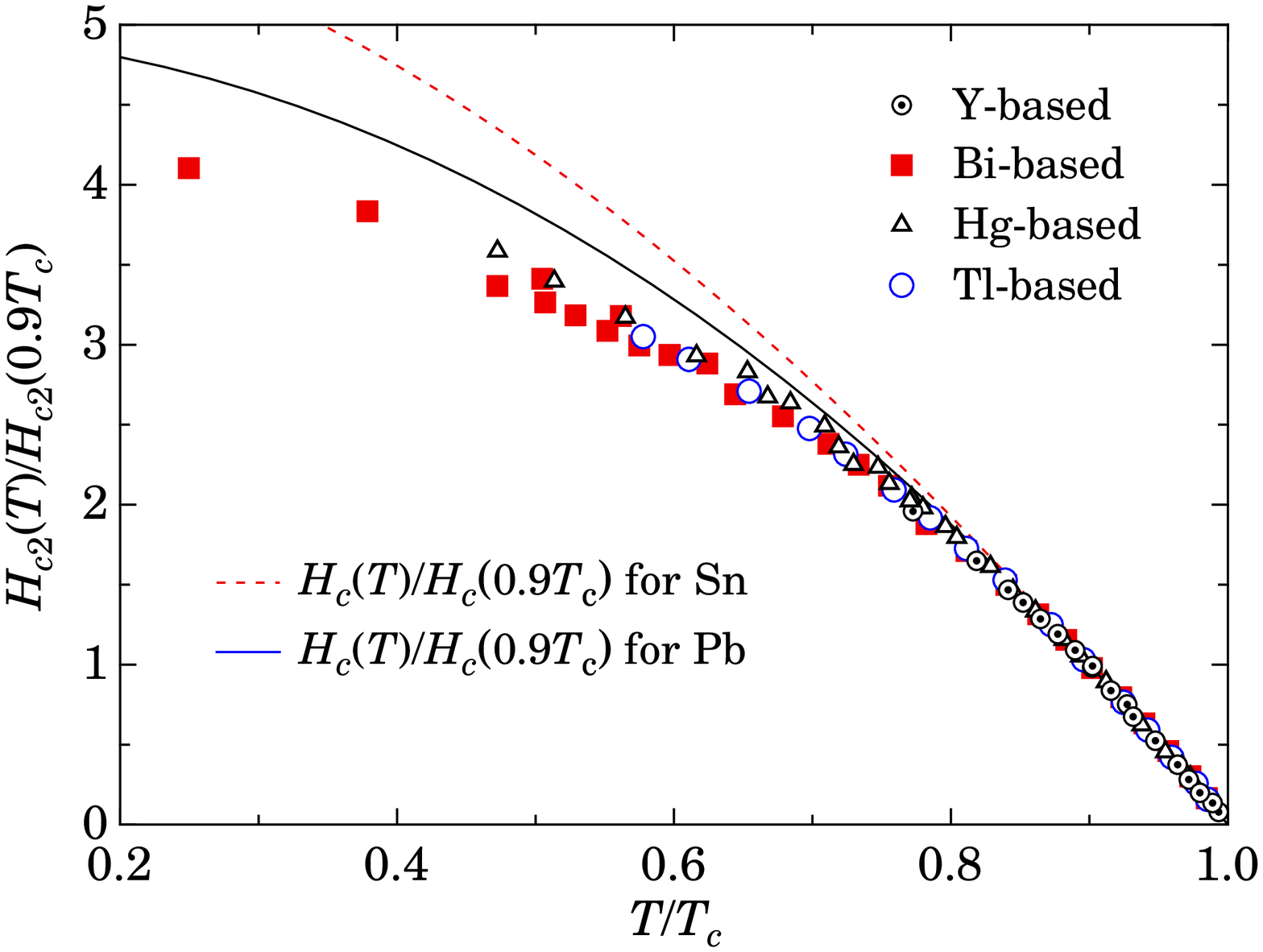,height=2 in}}
\caption{The normalized temperature dependence of $H_{c2}$ for different 
         HTSC compounds. Only some selected data points from Figs. 
         12(a-d) are shown. The solid and the broken line represent the 
         ratios $H_{c}(T)/H_{c}(0.9T_{c})$ for pure metallic Lead and 
         Tin, respectively.}  
\end{figure}
%%%%%%%%%%

The scaling procedure based on Eq. (17) turns out to be rather successful 
for the analysis of the reversible magnetization of HTSC's. The most surprising 
and completely unexpected result of our analysis is that for practically 
all families of HTSC's, the $h_{c2}(T/T_{c})$ curves are virtually identical. 
It is difficult to imagine that this universality of the $h_{c2}(T/T_{c})$ 
dependence is just a coincidence. We strongly believe that the spectacular 
agreement between the $h_{c2}(T/T_{c})$ data for a great variety of different 
samples is an unambiguous evidence that our approach captures the essential 
features of the magnetization process in HTSC's. It does not necessarily 
mean, however, that the Ginzburg-Landau parameter $\kappa$ is indeed 
temperature independent. The universality of $h_{c2}(T/T_{c})$ is preserved 
if the temperature dependence of $\kappa$ is the same for the different 
HTSC compounds studied here. 

Our analysis is applicable only to reversible magnetization data and 
therefore, all the results and conclusions are limited to temperatures 
more or less close to $T_{c}$. The lower limit of validity is quite different 
for different families of HTSC'c as may be seen in Figs. 12(a-d). 

The universality of the normalized temperature dependence of $H_{c2}$ implies 
that the normalized temperature variations of the thermodynamic critical 
fields $H_{c}(T)$ for different HTSC's are also identical. Since 
$H_{c}^{2}/8 \pi$ is the difference in the free energy densities 
between the normal and superconducting states, $H_{c}(T)$ also reflects the 
temperature dependence of the energy gap $\Delta$. In other words, our result 
that the normalized temperature dependence of $H_{c2}$ follows the same 
universal curve for different families of HTSC's implies that the normalized 
temperature variations of the energy gap $\Delta (T)/\Delta (0)$ for different 
HTSC's are also identical.

We note that the temperature dependencies of $H_{c2}$ for HTSC's obtained as 
outlined above are qualitatively very similar to those for conventional 
superconductors. They are linear at temperatures close to $T_{c}$ with 
a pronounced negative curvature at lower temperatures. Apparently, the 
positive curvature of $H_{c2}(T)$ for HTSC's, which is often reported in 
the literature, is due to the uncertainty of the definition of $H_{c2}$ 
used in those studies. 

\section{ALTERNATIVE MODEL OF THE MIXED STATE OF TYPE-II SUPERCONDUCTORS 
         IN MAGNETIC FIELDS CLOSE TO $H_{c2}$}

It is commonly accepted that the mixed state of a type-II superconductor 
is characterized by the penetration of an external magnetic field into 
the sample along quantized vortex lines or Abrikosov vortices. In most 
cases the vortices may be considered as thin normal filaments embedded 
in a superconducting environment. A completely different situation may, 
however, be established in superconductors with a GL parameter 
$\kappa \gg 1$ in magnetic fields close to $H_{c2}$. In this case, the 
distance between adjacent vortex cores is much smaller than the magnetic 
field penetration depth and the density of shielding currents is negligibly 
small. Here we suggest that in magnetic fields close to $H_{c2}$, the 
natural alternative to the conventional vortex structure is the formation 
of superconducting filaments embedded in the matrix of the normal metal. 

It is easy to show that the main interaction between such filaments is 
a short-range repulsion.\cite{7} This is why the filaments should always 
be separated by regions where the superconducting order parameter $\psi$ 
is zero. This circumstance makes it possible to analyze the properties 
of such superconducting filaments by numerically solving the GL equations 
for a single filament with the condition that $\psi = 0$ along the boundary 
between the filaments. Subsequently, various characteristics of the mixed 
state consisting of superconducting filaments such as the density of the 
free energy, the diamagnetic response, and the equilibrium density of 
filaments may be evaluated. \cite{7} The corresponding period $D_{f}$ of 
a triangular structure of superconducting filaments is plotted in 
Fig.14(a) as a function of $(1- H/H_{c2})$ together with the same quantity 
$D_{v}$ for the vortex lattice. As may be seen, $D_{f} \gg D_{v}$. 
%%%%%%%%%%%
\begin{figure}[h]
\centerline{\psfig{file=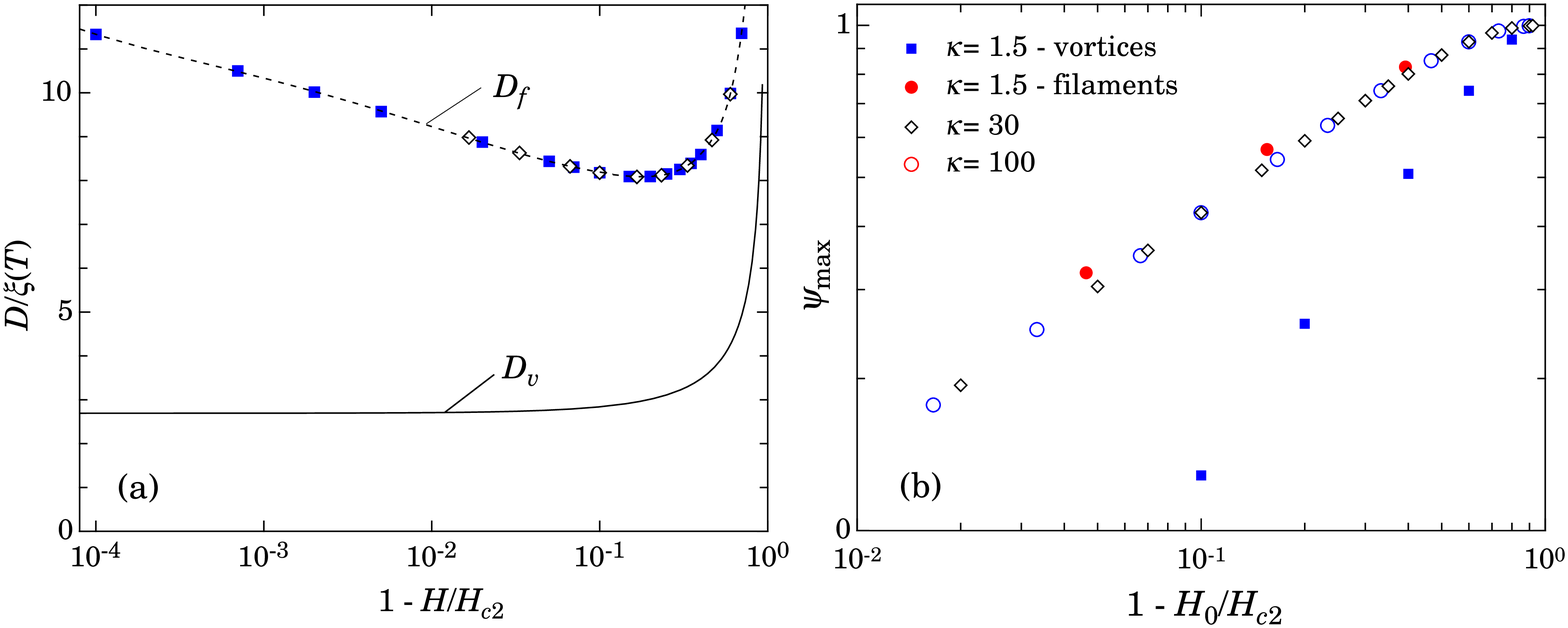,height=2 in}}
\caption{(a) Equilibrium periods of the system of superconducting filaments 
         $D_{f}$ and the vortex lattice $D_{v}$, respectively, as functions 
         of the applied magnetic field. (b) The maximum amplitude of the 
         order parameter for the system of filaments and for the vortex 
         lattice. The data for vortices are taken from Ref. 
         \protect\citen{58}.}  
\end{figure}
%%%%%%%%%%

Because each filament provides a negative contribution to the free energy, 
they should form a rather dense triangular configuration similar to that 
for a traditional vortex lattice. In order to use the 1-dimensional GL 
equations, we have to assume that the filaments are cylindrical. It is 
obvious, however, that, because of their mutual interaction, the filaments 
should adopt a hexagonal rather than a circular cross-section. This 
simplification is expected to lead to only slightly overestimated free 
energy values. 

In the high-magnetic-field limit, the GL free energy reduces to 
%%%%%%%
\begin{equation}
F=-{{H_c^2} \over {16\pi }}\int {\psi ^4(r)d^3r}
\end{equation}
%%%%%%%
with integration over the volume of the sample. \cite{34} Considering Eq. 
(18), it seems almost obvious that the vortex structure corresponds to 
a lower free energy than that for superconducting filaments. Indeed, in 
the case of the vortex lattice, the order parameter vanishes only along 
the vortex axes, while for the configuration of filaments the same happens 
along 2-dimensional inter-filament boundaries. In this case, the integration 
in Eq. (18) leads to a large numerical factor in favor of the vortex lattice. 
The actual situation, however, is more complex. Because each vortex line 
carries one magnetic flux quantum, the vortex density $n_{v}$ is strictly 
determined by the value of the applied magnetic field $H$. In the 
high-field limit, $n_{v} = H/\Phi_{0}$. There is no magnetic flux 
quantization condition for the system of superconducting filaments and 
therefore the density $n_{f}$ of filaments is a free parameter which may 
adjust itself to lower the free energy of the sample. In Fig.14(b) we show 
the field dependence of the order parameter amplitude $\psi_{\max}^{(f)}$ 
for the system of superconducting filaments in comparison with the same 
quantity $\psi_{\max}^{(v)}$ for the vortex lattice calculated in Ref. 
\citen{58}. In both cases $\psi_{\max}$ vanishes at $H = H_{c2}$, however, 
as may be seen in Fig.14(b), in the case of filaments, $\psi_{\max}^{(f)}$ 
is proportional to $\sqrt {1-H/H_{c2}}$, while for the vortex lattice, $\psi 
_{\max}^{(v)}\propto (1-H/H_{c2})$. This means that, in spite of the 
numerical factor mentioned above, the free energy of the system of filaments 
is lower than that for the vortex lattice in the limit of $H\to H_{c2}$.

The properties of a mixed state consisting of superconducting filaments 
are quite different from that which contains Abrikosov vortices. First, 
because the filaments are always separated by normal conducting regions, 
the sample resistance for currents perpendicular to the direction of the 
magnetic field never vanishes and the true zero-resistance superconducting 
state may be achieved only after a transition to the vortex structure. 
Second, in the case of filaments, the magnetic flux faces no barriers to 
move in or out of the sample and the magnetization of the sample must be 
reversible, independent of whether the filaments are pinned or not. With 
decreasing external magnetic field, the configuration of superconducting 
filaments necessarily has to undergo a transition to the conventional mixed 
state, involving Abrikosov vortices. The value of the transition field is 
determined by the free-energy balance between these two configurations which 
cannot be determined without more precise calculations of the free energy 
for both cases. The transition from one type of mixed state to the other 
involves a complete change of topology and must be accompanied by 
discontinuities in both the resistivity and the magnetic moment of the 
sample. We also expect some hysteresis, as well as a latent heat, dictated 
by the discontinuity of the magnetization. In other words, this transition 
is expected to exhibit all the features of a first-order phase transition. 
Similar transitions are observed in high-$T_{c}$ superconductors at $H < 
H_{c2}$ and are usually attributed to the melting of the vortex 
lattice. \cite{59,60,61,62,63}

As we have seen, in the high-magnetic field range, the distance between 
filaments is several times larger than the separation of vortices in the 
vortex lattice (Fig. 14(a)). This difference can be the key element to 
distinguish between these two realizations of the mixed state experimentally. 
This is not a simple task, however. In the high-magnetic-field range, where 
the mixed state consisting of superconducting filaments is expected to 
exist, the magnetic field is distributed almost uniformly across the 
sample and experiments that might distinguish between the mentioned 
options are quite difficult.

\end{document}